# Distinct magnetic regimes through site-selective atom substitution in the frustrated quantum antiferromagnet $Cs_2CuCl_{4-x}Br_x$

[1]P. T. Cong, [1]B. Wolf, [1]M. de Souza, [1]N. Krüger, [1]A. A. Haghighirad, [4]S. Gottlieb-Schoenmeyer, [1]F. Ritter, [1]W. Assmus, [2]I. Opahle, [2]K. Foyevtsova, [2]H. O. Jeschke, [2]R. Valentí, [3]L. Wiehl, and [1]M. Lang

[1]Physikalisches Institut, Goethe-Universität Frankfurt, SFB/TR 49, Max-von-Laue-Strasse 1, D-60438 Frankfurt(M)

[2]Institut für Theoretische Physik, Goethe-Universität Frankfurt, SFB/TR-49, Max-von-Laue-Strasse 1, D-60438 Frankfurt(M)

[3]Institut für Geowissenschaften, Goethe-Universität Frankfurt, Altenhöferallee 1, D-60438 Frankfurt(M)

[4]Physik-Department E21, Technische Universität München, James-Franck-Strasse 1,

D-85748 Garching

**Abstract**

We report on a systematic study of the magnetic properties on single crystals of the solid solution $Cs_2CuCl_{4-x}Br_x$ ($0 \leq x \leq 4$), which include the two known end-member compounds $Cs_2CuCl_4$ and $Cs_2CuBr_4$, classified as quasi-two-dimensional quantum antiferromagnets with different degrees of magnetic frustration. By comparative measurements of the magnetic susceptibility $\chi(T)$ on as many as eighteen different Br concentrations, we found that the in-plane and out-of-plane magnetic correlations, probed by the position and height of a maximum in the magnetic susceptibility, respectively, do not show a smooth variation with x. Instead three distinct concentration regimes can be identified, which are separated by critical concentrations $x_{c1} = 1$ and $x_{c2} = 2$. This unusual magnetic behavior can be explained by considering the structural peculiarities of the materials, especially the distorted Cu-halide tetrahedra, which support a site-selective replacement of $Cl^-$ by $Br^-$ ions. Consequently, the critical concentrations $x_{c1}$ ($x_{c2}$) mark particularly interesting systems, where one (two) halide-sublattice positions are fully occupied.

**Introduction**

Low-dimensional (low-D) quantum magnets reveal a wealth of fascinating and unexpected phenomena. Of particular interest are the anomalous properties which result from the interplay of strong quantum fluctuations and geometric frustration [1, 2, 3]. The simplest model for such a scenario is the spin S = ½ 2D triangular-lattice Heisenberg antiferromagnet [4, 5, 6]. A good realization of the spatially anisotropic version of this model is provided by the isostructural layered compounds $Cs_2CuCl_4$ [7, 8] and $Cs_2CuBr_4$ [9]. For $Cs_2CuCl_4$, the frustration effects are derived from a dominant antiferromagnetic exchange coupling $J/k_B$ = 4.34(6) K [10] along the in-plane *b*-axis together with a second in-plane coupling $J' \sim J/3$ along a diagonal bond in the *bc*-plane [11]. Further couplings in this material include a small inter-plane interaction $J'' \sim J/20$ as well as an anisotropic Dzyaloshinskii-Moriya interaction $D \sim J/20$ [11]. This material has attracted much attention due to its spin-liquid properties [7, 12] and its field-induced quantum phase transition around $B_s \sim 8.5$ T, separating long-range antiferromagnetic order below $T_N$ and $B \leq B_s$ from a fully polarized ferromagnetic state at $B > B_s$. The analogy of the critical properties to that of Bose-Einstein condensation has been pointed out [13], suggesting that it is the kinetic energy of delocalization of magnetic triplet excitations, which governs the physics in $Cs_2CuCl_4$ near $B_s$ [14].

On the other hand for $Cs_2CuBr_4$, where Néel ordering occurs at $T_N$ = 1.42 K, the magnetization shows a field-induced plateau at about one-third of the saturation magnetization [9]. The latter observation indicates that here, frustration effects are more pronounced, giving rise to the localization of the triplet excitations [14]. These border cases thus motivate the study of the magnetic properties of the solid solution $Cs_2CuCl_{4-x}Br_x$, in which by a continuous replacement of Cl⁻ by Br⁻, frustration effects are expected to become increasingly important. Much to our surprise, however, we find a discontinuous evolution of the magnetic properties of $Cs_2CuCl_{4-x}Br_x$ with x. Our study, which covers as many as 18 different Br concentrations, reveal three distinct magnetic regimes, with border-line concentrations $x_{c1}$ = 1 and $x_{c2}$ = 2 marking particularly interesting cases. We argue that the different character of these three phases can be understood on the basis of the materials' structure which supports a site-selective replacement of the halide ions.

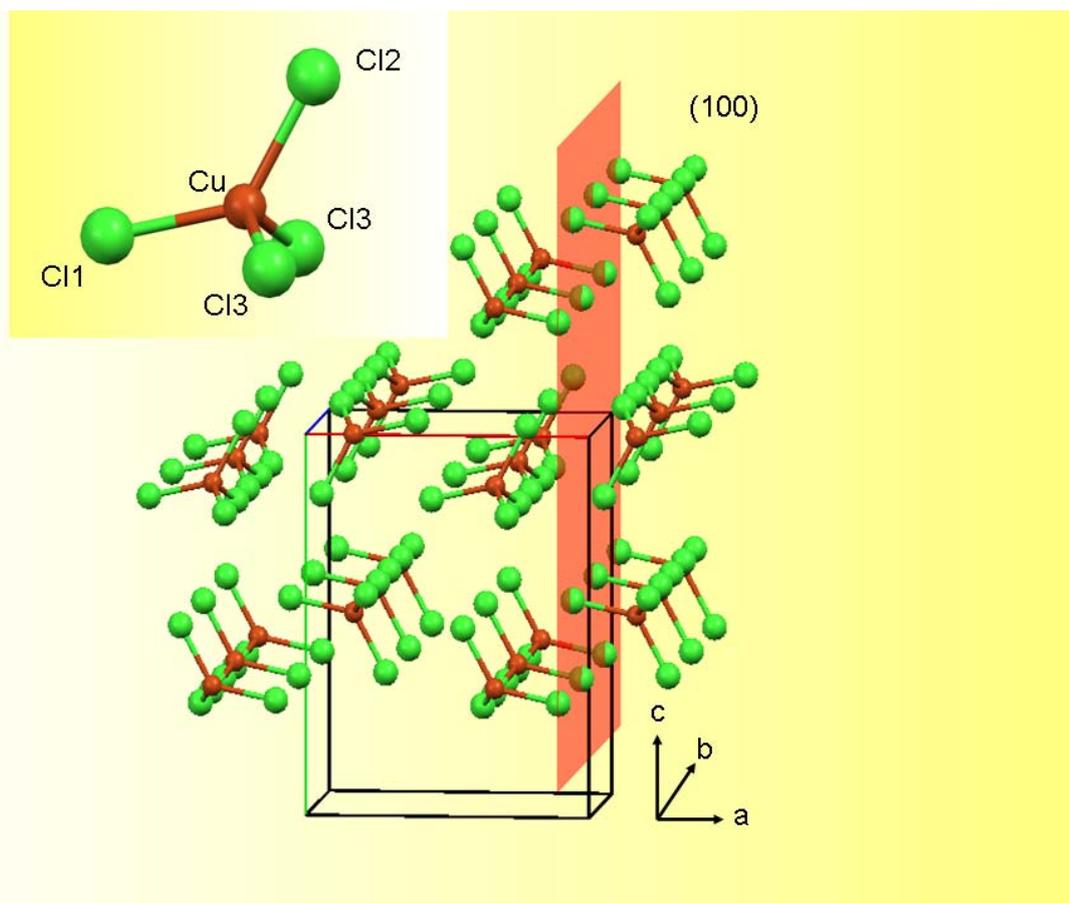

**Figure 1:** View with a tilt of 6° along the *a*-axis and rotated by - 6° around the *c*-axis of the orthorhombic structure of $Cs_2CuCl_4$. Cs-atoms are omitted for clarity. The longest Cu-Cl1 bonds are pointing along the *a*-direction perpendicular to the magnetic layers which are parallel to the (100) plane shown in red. **Inset:** Strongly flattened Cu-Cl tetrahedron with the three unequal Cl bonds. The bond lengths are listed in table 1.

The structure of $Cs_2CuCl_4$ at ambient conditions was determined by several authors as orthorhombic with space group *Pnma* [15, 16]. In Ref. [17] space group *Pnam*, a different setting of *Pnma*, was used. For the isostructural compound $Cs_2CuBr_4$ the space group was assigned to *Pnma* [18, 19, 20]. In both border-case compounds, the $Cu^{2+}$-ions are four-fold coordinated by halide ($Cl^-/Br^-$) ions, forming strongly flattened tetrahedra, cf. Fig. 1. These tetrahedra are well isolated from each other as they do not share any common coordination element. The shortest Cl-Cl distance between two adjacent tetrahedra is along the *b*-axis and amounts to 3.634 Å, significantly larger than the covalent radius of chlorine. All other distances between Cl-atoms belonging to adjacent tetrahedra are even longer. Adjacent planes are separated from each other by $Cs^+$-ions (not shown in figure 1) resulting in a quasi-2D spin $S = \frac{1}{2}$ arrangement.

An important structural feature, shared by both compounds, originates from the local Cu environment. As a consequence of the flattened Cu-halide tetrahedron, the bond lengths are significantly different. There is a longest Cu-halide bond, Cu-Cl1/Br1, and two equivalent shortest Cu-Cl3/Br3 bonds, cf. inset of Fig. 1 and table 1 for structural data taken from refs. [17, 20].

|  | Cu-Cl bond length [Å] | Bond length / Covalent radius |  | Cu-Br bond length [Å] | Bond length / Covalent radius |
|---|---|---|---|---|---|
| **Cl 1** | 2.244 | 0.959 | **Br 1** | 2.385 | 0.946 |
| **Cl 2** | 2.235 | 0.956 | **Br 2** | 2.362 | 0.938 |
| **Cl 3** | 2.220 | 0.940 | **Br 3** | 2.346 | 0.931 |

**Table 1:** Bond lengths of the flattened Cu-halide tetrahedron in $Cs_2CuCl_4$ and $Cs_2CuBr_4$ taken from refs. [17, 20] together with the ratio of the bond length to covalent radius.

The tetrahedra are oriented such that the long Cu-Cl1/Br1 bonds point along the inter-layer *a*-axis. As a result, this bond is involved in mediating the inter-layer exchange $J''$. In contrast, the Cu-Cl3/Br3 bonds are located within the *bc*-plane and are oriented along the *b*-axis, where adjacent Cu-halide tetrahedra have the shortest distance. This suggests that as a consequence the Cu-Cl3/Br3 bonds are involved in mediating the dominant magnetic interaction *J*.

**Experimental and computational details**

Large (100 – 1000 mm$^3$) and high-quality single crystals of the Cs$_2$CuCl$_{4-x}$Br$_x$ (0 ≤ $x$ ≤ 4) mixed system were grown from aqueous solutions by an evaporation technique, see ref. [21] for details. This method is preferred over simple temperature reduction for initiating and maintaining the growth process, because, depending on the temperature during the crystal growth, either an orthorhombic or a tetragonal structure can be realized for a given composition, especially for the concentration range ~1 ≤ $x$ ≤ ~2 and temperatures around 300 K [21]. For the latter concentration range, the orthorhombic structure is only obtained if the temperature of the solutions is kept above 323 K. In addition, to ensure that only crystals of the orthorhombic structure type were used for the present investigations, the crystals in the intermediate concentration range were heated up to above 423 K for a short time and then cooled down slowly to room temperature prior to the magnetic measurements. The compositional analysis of the various samples was conducted via EDX implying error bars for the concentration of δx = ± 0.075 [21].

The magnetic susceptibility was measured in the temperature range between 2 K ≤ $T$ ≤ 300 K and in magnetic fields of $B$ = 0.1 T and $B$ = 1 T using a Quantum Design SQUID magnetometer. All measurements were performed on single crystals grown in the orthorhombic phase with the external field oriented parallel to the crystallographic *b*-axis. The data were corrected for the temperature-independent diamagnetic core contribution, according to ref. [22], and the magnetic contribution of the sample holder. The latter was determined from an independent measurement. A careful determination of the magnetic background is important, especially at high temperatures where the materials' magnetic signal is small.

In addition, density functional theory (DFT) calculations were carried out to estimate the preferential position of the Br atoms in the mixed ($x$ = 1) compound Cs$_2$CuCl$_3$Br. The calculations were performed with the full potential local orbital code (FPLO, version 9.00-34) [23] using the experimental lattice parameters and atom positions [21] as input. The local density approximation (LDA) in the parameterization of Perdew and Wang [24] as well as the generalized gradient approximation (GGA) [25] were used to ensure that our conclusions do not depend on the choice of functional approximation to DFT. The calculations were performed in the scalar relativistic approximation with 216 **k**-points in the full Brillouin zone.

For a single crystal in the concentration range $1 \leq x \leq 2$ ($Cs_2CuCl_{2.7}Br_{1.3}$), X-ray intensity data at 103 K were collected using an Xcalibur3 four-circle diffractometer from Oxford Diffraction equipped with a Sapphire3 CCD camera and a sealed tube with MoKα radiation.

**Results and Discussion**

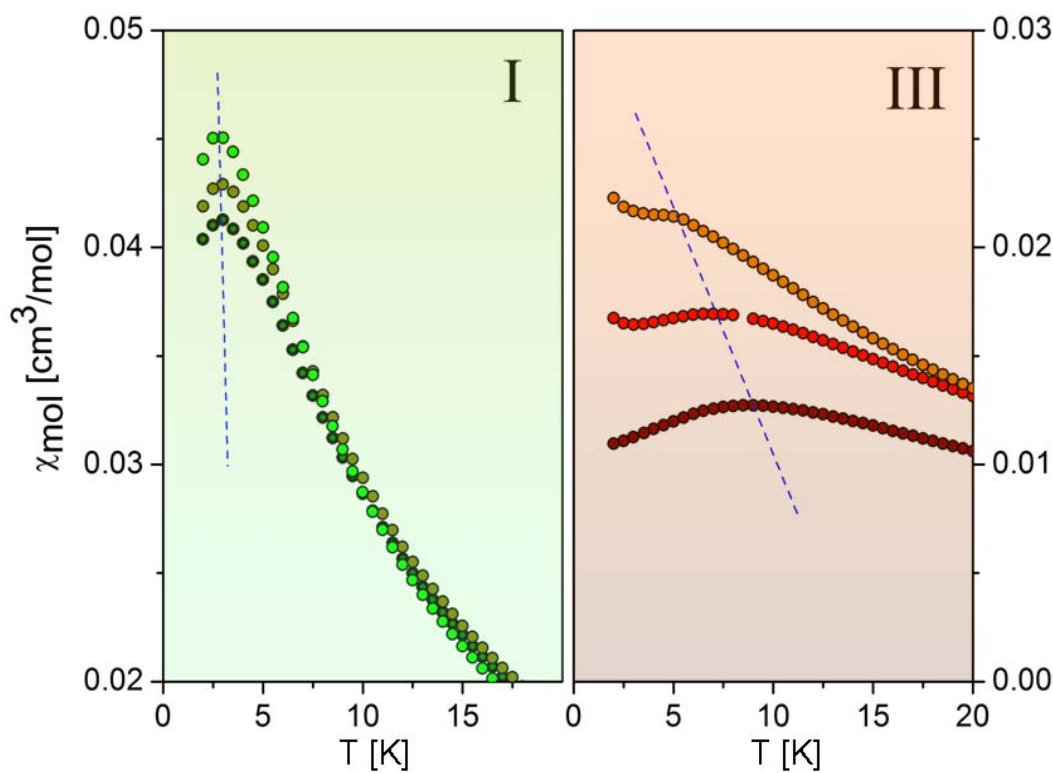

**Figure 2:** Overview of the molar magnetic susceptibility of the mixed system $Cs_2CuCl_{4-x}Br_x$ ($0 \leq x \leq 4$) for selected Br concentrations measured at $B = 0.1$ T. Left panel: Cl-rich side regime **I**: x = 0 (light green), x = 0.6 (dark yellow) and x = 0.8 (dark green); right panel: Br-rich side regime **III**: x = 2.2 (orange), x = 3.2 (red) and x = 4 (brown). The broken blue lines are guides to the eye indicating the evolution of $T_{max}(x)$. The color code, green for the Cl-rich side and brown for the Br-rich side, is used throughout the paper.

Figure 2 shows a compilation of the results of the molar susceptibility $\chi_{mol}(T)$ taken at a field of 0.1 T of the $Cs_2CuCl_{4-x}Br_x$ ($0 < x < 4$) mixed system for temperatures $2\,K \leq T \leq 20\,K$ [26]. The figure also includes the susceptibility data for the border cases $Cs_2CuCl_4$ (left panel, full light green circles) and $Cs_2CuBr_4$ (right panel, brown full circles), which are in accordance with literature results [27, 9, 28]. The susceptibility curves of these two compounds reveal a continuous increase with decreasing temperature and a maximum at $T_{max} = (2.8 \pm 0.15)$ K for $Cs_2CuCl_4$ and $(8.8 \pm 0.1)$ K for $Cs_2CuBr_4$, which is distinctly broader for the latter compound due to the larger J'/J ratio of 0.74 [28] as compared to J'/J = 0.34 for $Cs_2CuCl_4$ [11]. The maximum reflects the low-dimensional magnetic character of both materials in the temperature range under investigation. The occurrence of long-range magnetic order at $T_N = 0.62$ K ($Cs_2CuCl_4$) [10] and 1.42 K ($Cs_2CuBr_4$) [8], lying outside the accessible temperature range here, demonstrates the presence of weak 3D magnetic interactions.

A parameterization of the data in Fig. 2 and Fig. 3 with respect to the position of the susceptibility maximum and its height, $T_{max}$ and $\chi_{mol}(T_{max})$, respectively, shows that there is no continuous evolution of the magnetic properties as a function of the Br concentration. Instead, three distinct concentration regimes can be identified. The left panel of Fig. 2 exhibits data for small Br concentrations, i.e. $x \leq 1$, which will be labeled **regime I** in the following. Here the temperature dependence of $\chi_{mol}(T)$, including the shape and width of the maximum, is very similar to that of the pure chlorine system x = 0. In fact, for $Cs_2CuCl_{3.2}Br_{0.8}$, $T_{max}$ of $(3.0 \pm 0.12)$ K is very close to $T_{max}$ of the x = 0 compound, given the experimental uncertainties. With $\chi_{mol}(T_{max}) = 0.041$ cm$^3$/mol for x = 0.8, the value is reduced by slightly more than 10% compared to the pure chlorine (x = 0) system. This reduction of $\chi_{mol}(T_{max})$ is distinctly larger than the experimental error bars. Within **regime I**, some variations in the magnetic properties with x become noticeable in the high-temperature tail of the susceptibility, i.e., for temperatures 10 K ( ~ 2J ) $\leq T \leq$ 300 K. Here the data can be well described by a Curie-Weiss-type susceptibility $\chi_{cw} = C \cdot (T - \Theta_W)^{-1}$, with C the Curie constant and $\Theta_W$ the antiferromagnetic Weiss temperature, ranging from $\Theta_W = -(3.5 \pm 0.1)$ K ($x = 0$) to $\Theta_W = -(5.7 \pm 0.15)$ K ($x = 0.8$).

However, for the Br-rich side, i.e., x = 2 – 4, labeled **regime III** (right panel of figure 2), a change of the Br-content by the same amount of Δx = 0.8 has a much stronger effect on $T_{max}$ and $\chi_{mol}(T_{max})$. For $Cs_2CuCl_{0.8}Br_{3.2}$ (red full circles in the right panel of Fig. 2), for example, the data reveal a pronounced shift of $T_{max}$ to (6.9 ± 0.1) K and an accompanied strong increase of $\chi_{mol}(T_{max})$ by about 33% compared to the pure bromine system x = 4. Similar to **regime I**, the Weiss temperature changes markedly with x from $\Theta_W$ = -(18.4 ± 0.1) K ($x$ = 4) to $\Theta_W$ = -(9.8 ± 0.1) K ($x$ = 1.9). A peculiarity of the data for x = 3.2 in **regime III** is an upturn below about 3 K which grows with decreasing x and can be followed up to the border to **regime II**. Further experiments, in particular magnetic measurements at lower temperatures, have to clarify the origin of this feature.

The **regimes I** and **III,** characterized by distinctly different slopes of $T_{max}(x)$, reflect a continuous evolution of the magnetic properties of the corresponding end-member compounds x = 0 and x = 4, respectively, whereas crystals with $1 \leq x \leq 2$, labeled **regime II** henceforth, show a somewhat different magnetic behavior as displayed in figure 3. The data in figure 3 are taken in steps of Δx = 0.2. The Br concentration x = 0.8 in **regime I** and x = 2.2 in **regime III** are shown as dashed lines to highlight the difference in the magnetic properties in between the regimes. In **regime II** $\chi_{mol}(T_{max})$ decreases more rapidly with x compared to the reductions revealed in **regimes I** and **III**, cf. also Fig. 4. In addition, the data suggest a discontinuous change of $\chi_{mol}(T_{max})$ upon entering **regime I**. A less clear situation is encountered at the border to **regime III**. With decreasing the Br-content to about half way of the Br-Cl concentration range in $Cs_2CuCl_{1.8}Br_{2.2}$ (dark blue solid diamonds in Fig. 3), the characteristics of the Br-rich materials (**regime III)** are nearly preserved. The shift of $T_{max}$ to lower temperatures is accompanied by a further increase of $\chi_{mol}(T_{max})$. Note, that the broad maximum in the susceptibility is difficult to follow with increasing Br concentration due to the increase of $\chi_{mol}(T)$ below 4 K. As a consequence, significantly enlarged error bars have to be accepted rendering the boundary between **regime II** and **III** less well defined.

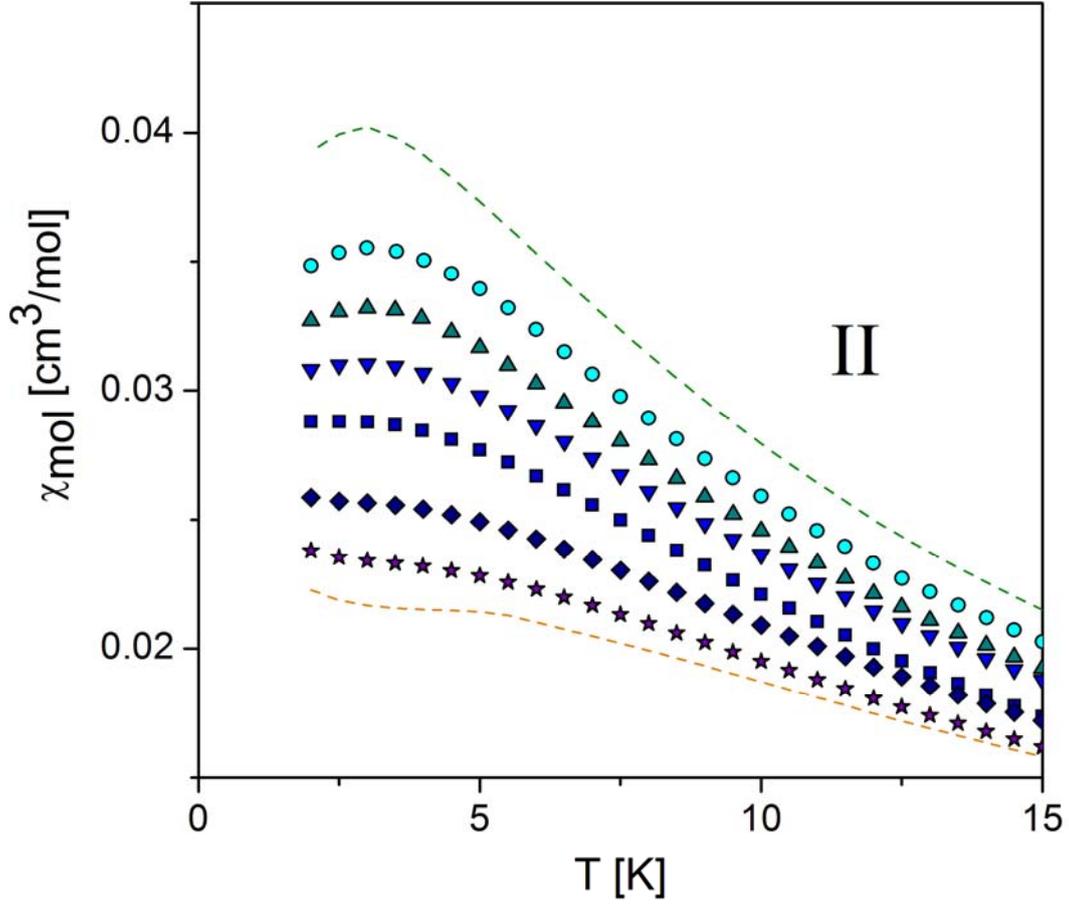

**Figure 3:** $\chi_{mol}(T)$ in **regime II** for various Br concentrations x = 1.0 (full cyan circles) to x = 2.0 (full violet stars) in steps of $\Delta x = 0.2$. The dashed lines show the $\chi_{mol}(T)$ data of x = 0.8 (green) in **regime I** and x = 2.2 (brown) in **regime III**.

The three distinct magnetic regimes in the $Cs_2CuCl_{4-x}Br_x$ ($0 \leq x \leq 4$) mixed system become obvious from Fig. 4, where the values for $T_{max}$ (blue solid squares) and $\chi_{mol}(T_{max})$ (red solid squares) are shown as a function of Br content x for all crystals investigated. On the chlorine rich side in **regime I** and up to $x$ slightly larger than 1.4, $T_{max}$ is nearly independent of the Br-concentration, while $\chi_{mol}(T_{max})$ reveals a distinct reduction with x. Note that for $x \geq 1.2$ the maximum in the magnetic susceptibility broadens and becomes less clearly pronounced as compared to that for $x \leq 1$. The fact that the characteristic temperature $T_{max}$ stays nearly constant indicates that the Br substitution in this concentration range leaves the quasi-2D magnetic fluctuations, caused by in-plane interactions, practically unaffected. At the same time, the significant reduction in $\chi_{mol}(T_{max})$ with increasing x demonstrates that the magnetic coupling between the layers is substantially modified. This is consistent with the results of exact diagonalization calculations for $Cs_2CuCl_4$, where the influence of the strength of the various interlayer couplings on the magnetic properties was investigated [29]. The authors found a decrease in $\chi_{mol}(T_{max})$ when the largest interlayer coupling $J3$ is included in the calculation without any shift of $T_{max}$. In contrast, for the Br-rich side (**regime III**), the strong decrease of $T_{max}$ by approximately a factor of 2 on decreasing x from 4 to 2 clearly signals the suppression of the 2D-magnetic correlations in this part of the concentration range.

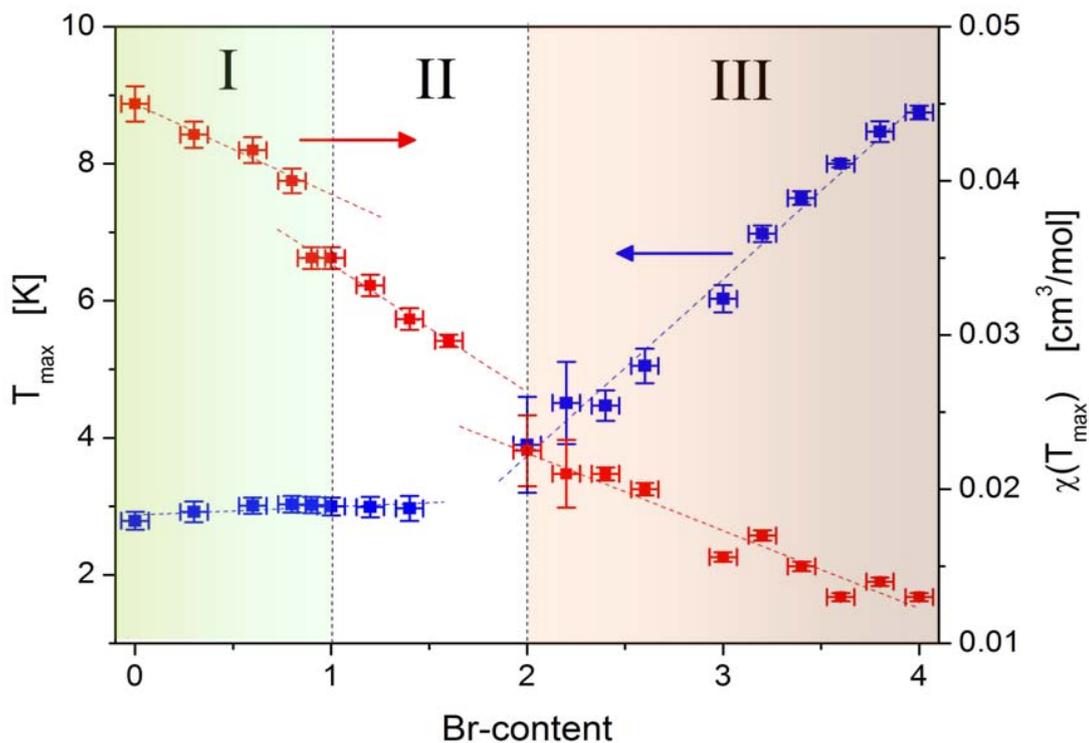

**Figure 4:** $T_{max}$ (blue solid squares, left axis) and $\chi_{mol}(T_{max})$ (red solid squares, right axis) as a function of the Br content x for all samples under investigation. The vertical dotted lines indicate the critical concentrations $x_{c1}$ = 1 and $x_{c2}$ = 2 separating the regimes I, II and III. The broken lines are linear fits to the experimental data in the three regimes. Green (brown) area denotes the Cl (Br)-rich side of the concentration range.

The magnetic properties of the $Cs_2CuCl_{4-x}Br_x$ ($0 \leq x \leq 4$) mixed systems and the corresponding three distinct magnetic regimes can be understood using the following model which considers structural features of the Cu-halide tetrahedra embedded in the orthorhombic crystal structure. Upon replacing the Cl⁻ ions by Br⁻ on the chlorine-rich side (**regime I**), the Cu-Cl1-bonds will be almost exclusively affected, as they exhibit the largest bond length and the Br⁻ ions are considerably bigger than the Cl⁻ ions. Since the Cu-Cl1-bonds point along the inter-layer *a*-axis of the orthorhombic structure, they mediate the magnetic exchange $J''$ between the layers. Thus, starting from the pure $Cs_2CuCl_4$ compound, the progressive substitution of the Cl1 atoms by Br will be accompanied by an increase of $J''$, consistent with the decrease of the magnetic susceptibility at its maximum. This site-selective replacement of the Cl1 ions in regime I, which for a hypothetically ideal system would be completed at a concentration $x_{c1} = 1$, also explains that $T_{max}$ is constant on the chlorine-rich side of the phase diagram: the 2D magnetic correlations, probed by $T_{max}$, involve the Cl2/Cl3 atoms, the occupation factors of which remain (practically) unaffected for Br-substitutions in **regime I** for $x \leq x_{c1}$. It should be mentioned that due to entropic reasons, the site-selective substitution will not be perfect, resulting in a finite disorder in the occupancy of the halide positions.

A quite different situation is encountered on the Br-rich side. Starting from the border case $Cs_2CuBr_4$ and substituting Br⁻ by the smaller Cl⁻ will predominantly affect the two equivalent Br3 positions with the shortest bond length. As a consequence, the intra-layer coupling constants $J$ and $J'$, i.e. the 2D magnetic correlations, will be reduced. This mechanism explains the strong reduction of $T_{max}$ as a function of Br content observed in the susceptibility measurement.

Further support for the above reasoning comes from our DFT calculations, which show a clear preference for the substitution of the Cl1 atoms by Br on the Cl-rich side: the occupation of the Cl1 position in $Cs_2CuCl_3Br$ yields the lowest total energy, whereas the energy of a structure where Br is substituted for Cl2 (Cl3) is about 222 (247) meV/Br higher in energy in the LDA calculations. In the GGA calculations, the corresponding energy differences are 244 (274) meV/Br, showing only a weak dependence on the choice of functional. These calculations indicate that the sterical aspects, i.e. ionic radii and corresponding bond lengths, covered by the above simple model, are of crucial importance for stabilizing the structure in the $Cs_2CuCl_{4-x}Br_x$ ($0 \leq x \leq 4$) mixed system. The reason for that lies in the peculiarity of the Cu-halide tetrahedron [30], which forms a stable, discrete structural unit, typically found in aqueous solutions, and which is also only weakly bond in the orthorhombic crystal structure of the $Cs_2CuCl_{4-x}Br_x$ ($0 \leq x \leq 4$) mixed system.

Another independent proof of this model based on the distribution of bond lengths inside the flattened Cu-halide tetrahedron is given by the relative length changes $[l(x)-l(x_0)] / l(x_0)$ of the lattice constants as a function of Br-content. As shown in ref. [21] $[l(x)-l(x_0)] / l(x_0)$ is isotropic in the concentration range $1 \leq x \leq 2$ and anisotropic in the regimes II and III. With increasing x for $0 \leq x \leq 1$, the relative expansion of the *a*-axis is largest, consistent with a predominant substitution of the Cl1 positions by Br whereas on the Br-rich side $2 \leq x \leq 4$, the Cl doping on the two Br3 positions leads to the strongest reduction of the *b*-axis.

While the above-mentioned results, derived from experiments and calculations, provide clear arguments in favor of a site-selective Br/Cl substitution, some quantitative evidence can be deduced from a preliminary X-ray crystal structure determination. The measurement was intended to ensure the orthorhombic structure of a single crystal with the composition $Cs_2CuCl_{2.7}Br_{1.3}$ belonging to **regime II**. The following lattice parameters were found at 103 K: *a* = 9.8821(2) Å, *b* = 7.6044(2) Å, *c* = 12.4746(3) Å ($\alpha = \beta = \gamma = 90°$) in space group *Pnma*. The structure was refined with SHELXL97-2, resulting in R1 = 2.39 % and wR2 = 6.24 %. The occupation factors of Br/Cl on the three symmetrical inequivalent sites were refined to 69.6(5) % Br on site 1, 25.3(5) % Br on site 2 and 15.6(4) % Br on site 3. Taking into account the respective site symmetries (no. 1 and 2 on the mirrorplane, no. 3 in general position) the overall Br : Cl ratio sums up to 1.26(2) : 2.74(2). Further details of the structural properties, especially for systems of the Br concentration range $1 \leq x \leq 2$ will be given in a separate paper.

**Summary and outlook**

We have performed measurements of the magnetic susceptibility on single crystals of the Cs$_2$CuCl$_{4-x}$Br$_x$ ($0 \leq x \leq 4$) mixed system in the orthorhombic phase. By following characteristic features of the susceptibility, including the position of the broad maximum at $T_{max}$ and the height of the maximum $\chi_{mol}(T_{max})$, three distinct magnetic regimes have been identified. These regimes are separated by critical concentrations around $x_{c1} = 1$ and $x_{c2} = 2$. The main characteristics of the systems' magnetic behavior and the existence of the two critical concentrations could be explained by a simple model which considers the structural peculiarities of the Cu-halide tetrahedra and the way these building blocks are arranged in the crystal structure. According to this model, the substitution of the smaller Cl$^-$ ions by the larger Br$^-$ ions in the distorted Cu-halide tetrahedron enforces a site-selective occupation. This mechanism provides a natural explanation for the two critical concentrations: while at $x_{c1} = 1$, (practically) all the Cl1 positions are occupied by Br, the two equivalent Br3 atoms are displaced by Cl atoms at $x_{c2} = 2$. Thus our results suggest that Cs$_2$CuCl$_3$Br$_1$ and Cs$_2$CuCl$_2$Br$_2$ mark particularly interesting mixed systems with a well-ordered local Cu environment. We anticipate that for the Cs$_2$CuCl$_2$Br$_2$ system the site-selective substitution of the two Br3 atoms by chlorine, which would weaken the dominant magnetic exchange J while leaving J' unaffected, gives this material an even more frustrated character compared to the pure system Cs$_2$CuBr$_4$. Thus Cs$_2$CuCl$_2$Br$_2$, which is expected to show long-range antiferromagnetic order at $T_N$ below that of the pure Cl-system ($T_N(x = 0) = 0.62$ K), represents an interesting target material for studying the interplay of strong magnetic frustration and quantum criticality for $T_N \to 0$.